# Flashes in a Star Stream: Automated Classification of Astronomical Transient Events

S. G. Djorgovski, A. A. Mahabal, C. Donalek,
M. J. Graham, A. J. Drake
California Institute of Technology
Pasadena, CA 91125, USA
[george,aam,donalek,mjg,ajd]@astro.caltech.edu

B. Moghaddam, M. Turmon
Jet Propulsion Laboratory
California Institute of Technology
Pasadena, CA 91109, USA
[baback,turmon]@jpl.nasa.gov

*Abstract*—An automated, rapid classification of transient events detected in the modern synoptic sky surveys is essential for their scientific utility and effective follow-up using scarce resources. This presents some unusual challenges: the data are sparse, heterogeneous and incomplete; evolving in time; and most of the relevant information comes not from the data stream itself, but from a variety of archival data and contextual information (spatial, temporal, and multi-wavelength). We are exploring a variety of novel techniques, mostly Bayesian, to respond to these challenges, using the ongoing CRTS sky survey as a testbed. The current surveys are already overwhelming our ability to effectively follow all of the potentially interesting events, and these challenges will grow by orders of magnitude over the next decade as the more ambitious sky surveys get under way. While we focus on an application in a specific domain (astrophysics), these challenges are more broadly relevant for event or anomaly detection and knowledge discovery in massive data streams.

## I. Introduction

A new generation of scientific measurement systems (instruments or sensor networks) is generating exponentially growing data streams, now moving into the Petascale regime, that can enable significant new discoveries. In many cases, scientific goals include detection of phenomena where a rapid change occurs, that have to be identified, characterized, and possibly followed by new measurements. The requirement to perform the analysis rapidly and objectively, coupled with huge data rates, implies a need for an automated event detection, classification, and follow-up decision making.

This entails some special challenges beyond traditional automated classification approaches, which are usually done in some feature vector space, with an abundance of self-contained data derived from homogeneous measurements. Here, the input information is generally sparse and heterogeneous: there are only a few initial measurements, that differ from case to case, and having differing measurement errors; the contextual information is often essential, and yet difficult to capture and incorporate in the classification process; many sources of noise, instrumental glitches, etc., can masquerade as transient events in the data stream; new, heterogeneous data arrive, and the classification must be iterated dynamically. Requiring a high completeness (don't miss any interesting events) and low contamination (a few false alarms), and the need to complete the classification process and make an optimal decision about expending valuable follow-up resources (e.g., obtain additional measurements using a more powerful instrument at a certain cost) in (near) real time are substantial challenges that require some novel approaches.

While this situation arises in many application domains, it is well exemplified in the developing field of time domain astronomy, with telescope systems dedicated to discovery of moving objects, e.g., potentially hazardous, Earth-crossing asteroids [1,2,3], transient or explosive astrophysical phenomena, e.g., supernovae (SNe), γ-ray bursts (GRBs), detection of extrasolar planets through occultation and microlensing flares, and so on – each of them requiring rapid alerts and follow-up observations. The time domain is rapidly becoming one of the most exciting new research frontiers in astronomy [23,29], broadening substantially our understanding of the physical universe, and it may lead to a discovery of previously unknown phenomena [16,23,24].

The key to progress in time-domain astrophysics is the availability of substantial event data streams generated by panoramic digital synoptic sky surveys, coupled with a rapid follow-up of potentially interesting events (photometric, spectroscopic, and multi-wavelength). A number of synoptic astronomical surveys are already operating [see, e.g., 1,2,3,7,17,25,26,43], and much more ambitious enterprises are being planned [4,5,6], moving into the Petascale regime, with hundreds of thousands of transient events per night, implying a need for automated, robust processing and follow-up. There is also a growing number of autonomous robotic telescopes geared to discovery and follow-up of rare transient events. Essentially, *a new generation of scientific measurement systems is emerging* in astronomy, and many other fields: connected sensor networks which gather and analyze data automatically, and respond to the outcome of these measurements in the real-time, often redirecting the measurement process itself, and without human intervention. Thus, the application of machine learning and machine intelligence methods becomes a natural and integral part of the scientific discovery process.

We are developing a novel set of techniques and methodology for automated, real-time data analysis and discovery, operating on massive and heterogeneous data streams from robotic telescope sensor networks, fully integrated with the Virtual Observatory (VO) framework

[39,40,42]. The system under development incorporates machine learning (ML) elements for an iterative, dynamical classification of astronomical transient events, based on the initial detection measurements, archival information, and newly obtained follow-up measurements from robotic telescopes.

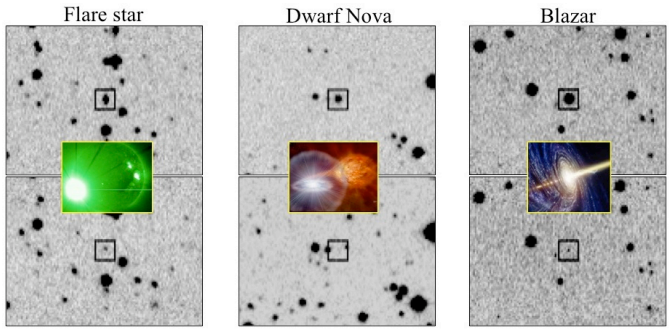

**Figure 1.** Examples of transient events from the Catalina Real-time Transient Survey (CRTS) [17,25]. Images in the top row show objects which appear much brighter that night, relative to the baseline images obtained earlier (bottom row). On this basis alone, the three transients are physically indistinguishable, yet the subsequent follow-up shows them to be three vastly different types of phenomena: a flare star (left), a cataclysmic variable powered by an accretion to a compact stellar remnant (middle), and a blazar, flaring due to instabilities in a relativistic jet (right). Accurate transient event classification is the key to their follow-up and physical understanding.

## II. THE CHALENGES OF AN AUTOMATED, REAL-TIME EVENT CLASSIFICATION

A full scientific exploitation and understanding of astrophysical events requires a rapid, multi-wavelength follow-up. The *essential enabling technologies* that need to be automated are robust classification and decision making for the optimal use of follow-up facilities. They are the key for exploiting the full scientific potential of the ongoing and forthcoming synoptic sky surveys. Some approaches to automated classification of astronomical sources, transients and variables include, e.g., [10, 11, 18, 19, 20, 27, 28, 30, 31, 41, 44, 45, 46, 47].

The goal is to associate classification probabilities that any given event belongs to a variety of known classes of variable astrophysical objects and to update such classifications as more data come in, until a scientifically justified convergence is reached [24]. The process has to be *as automated as possible, robust, and reliable*; it has to operate on *sparse and heterogeneous data*; it has to maintain a *high completeness* (not miss any interesting events) yet a *low false alarm rate*; and it has to *learn* from past experience for an ever improving, evolving performance.

The next step is a development and implementation of an automated follow-up event prioritization and decision making mechanism, which would actively determine and request follow-up observations on demand, driven by the event data analysis. This would include an automated identification of the most discriminating potential measurements from the available follow-up assets, taking into account their relative cost functions, in order to optimize both classification discrimination, and the potential scientific returns.

In the context of time-domain astronomy, we can distinguish two regimes regarding data mining and classification: (1) A non-time-critical, archival approach, where the relatively stable and extensive collection of data on some variable object or transient event can be used to classify it, and (2) Real-time classification of transient events detected in massive data streams, that can be used to guide immediate or time-critical follow-up observations. Generally speaking, regime (2) is much more challenging, both due to the time-critical nature, and the fact that often less information is available than in the regime (1); however, results and methods obtained in (1) can often be used in the real-time applications in (2). Here we focus on the latter.

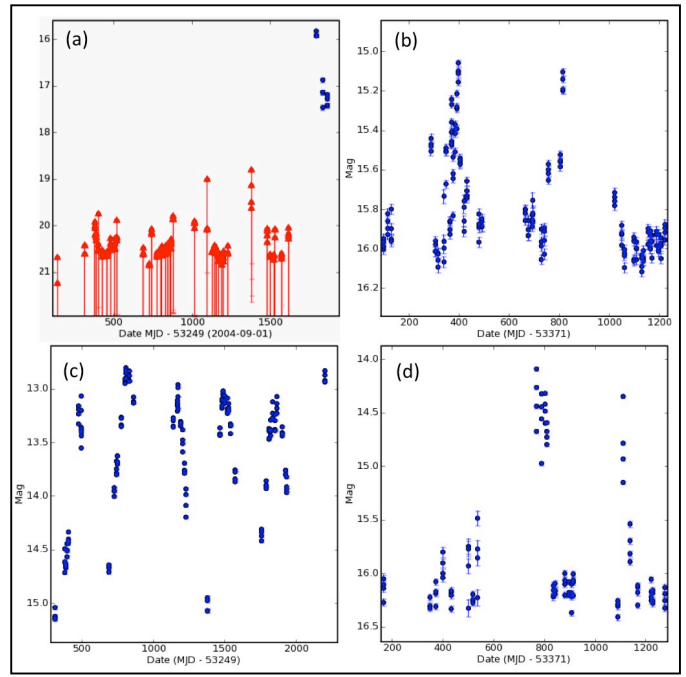

**Figure 2.** Examples of light curves (flux histories) for different types of variable astrophysical objects from the Catalina Real-time Transient Survey (CRTS) [17,25]: (a) Supernova (red triangles with down-pointing error bars are the upper limits, i.e., non-detections); (b) blazar (beamed active galactic nucleus); (c) pulsating variable star; (d) cataclysmic variable star. Such past light histories represent an essential temporal contextual information that can be used for event classification.

We can also distinguish between the *signal classification*, i.e., whether a detected event is real, or just an instrumental and/or a data processing artifact, and a *physical classification*, where we associate a given event with a range of different physical interpretations of its nature. In the context of synoptic sky surveys, the former problem has been successfully addressed by the use of Artificial Neural Networks (ANN) [50], and Support Vector Machines (SVM) [51], operating on feature vectors composed of morphological image parameters measured by the source detection pipeline. These automated approaches eliminate up to 95% of image artifacts, making the remaining filtering much more viable. We have also used contextual information, e.g., a strong spatial clustering of candidate events is almost always indicative of a spurious nature, using the a priori knowledge about the instrument or the scientific context.

The physical classification problem is much more difficult. The challenge here is that all genuine astrophysical transient events look the same in the images (PSF-like), so that information other than image morphology must be used. One problem is that in general, not all parameters would be measured for all events, e.g., some may be missing a measurement in a particular filter, due to a detector problem; some may be in the area on the sky where there are no useful radio observations; etc. This make the use of any feature-vector based classification methods, such as ANN, SVM, etc., very problematic, although we describe one approach to this methodology below. In general, we are driven towards methods that can operate on an arbitrary subset of measurements drawn from the continuous distributions of observables, such as the flux histories (light curves) in a particular bandpass.

Another problem is that many observables may be given as upper or lower limits, rather than as defined measurements; for example, "the increase in brightness is > 3.6 magnitudes", or "the radio to optical flux ratio of this source is < 0.01". One approach is to treat them as missing data, implying a loss of the potentially useful information. A better approach is to reason about "censored" observations, that can be naturally incorporated through a Bayesian model by choosing a likelihood function that rules out values violating the bounds.

### III. A BAYESIAN APPROACH TO EVENT CLASSIFICATION

The main astronomical inputs are in the form of observational and archival parameters for individual objects, which can be put into various, often independent subsets. Examples include various fluxes at different wavelengths, associated colors or hardness ratios, proximity values, shape measurements, magnitude characterizations at different timescales, etc. The heterogeneity and sparsity of data makes the use of Bayesian methods for classification a natural choice.

Distributions of such parameters need to be estimated for each type of variable astrophysical phenomena that we want to classify. Then an estimated probability of a new event belonging to any given class can be evaluated from all of such pieces of information available, as follows. Let us denote the feature vector of event parameters as $x$, and the object class that gave rise to this vector as $y$, $1 \leq y \leq K$. While certain fields within $x$ will generally be known, such as sky position and brightness in selected filters, many other parameters will be known only sporadically, e.g., brightness change over various time baselines. In a Bayesian approach, $x$ and $y$ are related via:

$$P(y = k \mid x) = P(x \mid y = k) P(k) / P(x) \propto$$
$$\propto P(k) P(x \mid y = k) \approx P(k) \prod_{b=1}^{B} P(x_b \mid y = k)$$

Because we are only interested in the above quantity as a function of $k$, we can drop factors that only depend on $x$. We assume that, conditional on the class $y$, the feature vector *decomposes* into $B$ roughly independent blocks, generically labeled $x_b$. These blocks may be singleton variables, or contain multiple variables, e.g., sets of filters that are highly correlated. The resulting algorithm is called *naive Bayes* because of its assumption that we may decouple the inputs in this way [8,9].

This decoupling is advantageous because it allows us to circumvent the "curse of dimensionality," because we will eventually have to learn the conditional distributions $P(x_b \mid y = k)$ for each $k$. As more components are added to $x_b$, more examples will be needed to learn the corresponding distribution. The decomposition keeps the dimensionality of each feature block manageable. Moreover, such decomposition allows us to cope easily with ignorance of missing variables: we simply drop the corresponding factors.

As a simple demonstration of the technique, we have been experimenting with a prototype Bayesian Network (BN) model [32,33]. We use a small but homogeneous data set involving colors of transients detected in the CRTS survey [17,25], as measured at the Palomar 1.5-m telescope. We have used multinomial nodes (discrete bins) for 3 colors, with provision for missing values, and a multinomial node for Galactic latitude which is always present and is a probabilistic indicator of whether an object is Galactic or not. The current priors used are for six distinct classes, cataclysmic variables (CVs), supernovae (SN), Blazars, other AGNs, UV Ceti stars and all else bundled into a sixth class, called Rest.

Using a *single* epoch measurement of colors, in the relative classification of CVs vs. SNe, we obtain a completeness of ~ 80% and a contamination of ~ 19%, which reflects a qualitative color difference between these two types of transients. In the relative classification of CVs vs. Blazars, we obtain a completeness of ~ 70 – 90% and a contamination of ~ 10 – 24%, which reflects the fact that colors of these two types of transients tend to be similar, and that some additional discriminative parameter is needed. We are currently expanding to a BN with an order of magnitude more classes, more observable parameters (e.g., flux and color measurements with more bandpasses and at different epochs), and additional BN layers.

In this framework the priors come from a set of observed parameters like distribution of colors, distribution of objects as a function of Galactic latitude, frequencies of different types of objects etc. The posteriors we are interested in are determining the type of an object based on, say, its (*r-i*) color, Galactic latitude and proximity to another object etc. Sparse and/or irregular light curves (LC) from any given object class can have sufficient salient structure that can be exploited by automated classification algorithms. We have experimented with Gaussian Process Regression [34], and found it to be useful for parameter estimation for a certain types of LCs that can be represented by a standard data model (e.g., Supernovae).

Due to the computational complexity of Gaussian Process modeling, we are also experimenting with a different and simpler probabilistic approach to characterizing LCs and their intrinsic shape and structure. By pooling a large ensemble of an object class's individual LCs (constituting a representative training set) we aim to model class-specific shape and structure probabilistically. We do this by constructing a histogram of empirical probability density function (PDF) over invariant LC descriptors or "signatures" (described below), which are then used for classification of new event observations (*test set*). Such comparisons (test probe *vs*. trained model) can be made either in batch mode (*e.g*, as offline searches in a LC database) as well as incrementally and on-line, as new observations "trickle in", with the revised classification scores growing more confident with each additional observation that is accumulated.

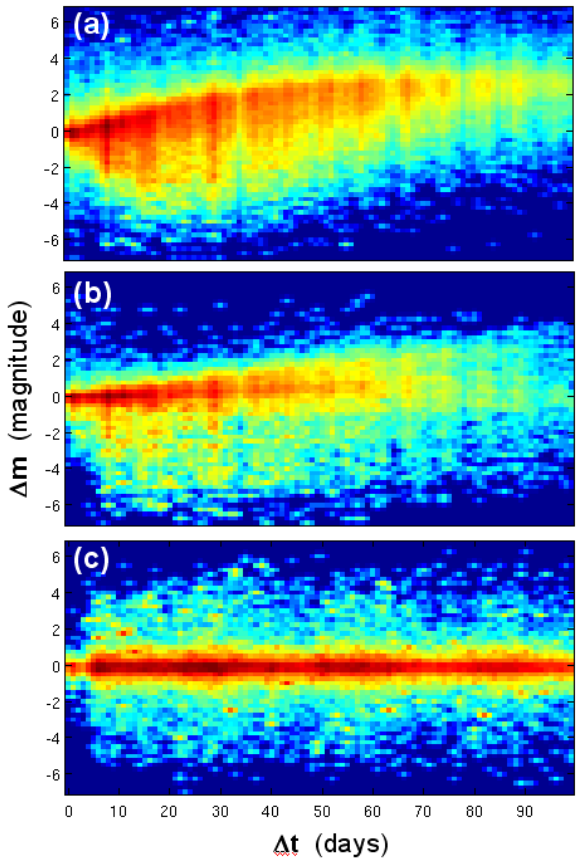

**Figure 3.** Probabilistic structure functions representing the joint distribution of ($\Delta t$, $\Delta m$) values from all ($\Delta t > 0$) paired observations in LCs, shown here as discretized 2D histograms of 3 classes of transients: (a) supernovae of type SN-Ia, (b) supernovae of type SN-IIp and (c) Cataclymic Variables, using bin widths of $\Delta t = 1$ day, and $\Delta m = 0.25$, smoothing with an anisotropic convolution kernel, and with pixel intensity corresponding to log-probability. These class prototype histograms were obtained by *pooling* {879, 282, 426} LCs from the corresponding 3 object classes. Note that the upward "arch" of the supernovae is due to their sustained flux decay (*increasing* $\Delta m$) and that the temporal/flux shape structure of all 3 classes forms a distinct signature. The probability of observed ($\Delta t$, $\Delta m$) values from a new (unknown) object's LC can therefore be easily "read off" (scored) by each histogram. Probabilistic structure functions can thus be viewed as "generative models" of ($\Delta t$, $\Delta m$) for their respective LC classes (*i.e.*, as nonparametric likelihood functions).

Since typical survey (flux-only) observations come in the form of magnitude changes over time increments – *($\Delta t$, $\Delta m$)* – we focus on modeling the joint distribution of all such pairs of values for a given LC (note: we consider all causal increments, corresponding to $\Delta t > 0$, therefore $n$ LC observations lead to $n(n-1)/2$ pairs). By virtue of being increments these *($\Delta t$, $\Delta m$)* change values and their PDF will be invariant to absolute magnitude and time as well as corresponding shifts in each (since distance to an object and "true" onset time of its LC are unknown). These densities allow flux upper limits to be encoded rather easily – e.g., under poor seeing conditions we may only have bounded observations, such as $m > 18$, which leads to a bounded $\Delta m$ (which maps to a vertical segment in the histogram, as opposed to a single bin).

We can also smooth our 2D histograms in order to model uncertainties in ($\Delta t$, $\Delta m$). Hence, this yields a computationally simple and effective way to implement a nonparametric density model that is flexible enough for the variety of object classes under consideration. Note that our histograms can be viewed as *probabilistic structure functions*: a standard s.f. simply gives flux variance (a scalar quantity) as a function of $\Delta t$, whereas here we have a full PDF on $\Delta m$, indexed by $\Delta t$ (from which a standard s.f. can be easily derived). Figure 3 shows examples of these 2D histograms for three classes of transient objects.

When a new transient is detected, its ($\Delta t, \Delta m$) histogram starts to be accumulated. After each new measurement, it is compared to a set of template histograms for different classes of transients. We apply a set of metrics that produces relative likelihoods of the new transient belonging to any given class. As the data accumulate, the classification accuracy improves.

In our preliminary experimental evaluations with few object classes (single-burst like SN, periodic variable stars like RR Lyrae / Miras, as well as stochastic sources like Blazars and CVs) we have found that these compact ($\Delta t$, $\Delta m$) density models have potential as a simple yet accurate classification framework, especially when faced with *sparse* and irregular monochromatic time series like typical observational LC data.

The next step in the development of this classifier is to use 4D histograms of data point triplets. For example, if we measure magnitudes $m_1$, $m_2$, and $m_3$ at times $t_1$, $t_2$, and $t_3$, the histogram axes are now ($\Delta t_{12}$, $\Delta m_{12}$, $\Delta t_{23}$, $\Delta m_{23}$). These 4D histograms are sparsely populated, but separate the different classes more clearly. This work is still in progress.

IV. CLASSIFICATION OF PARAMETRIZED LIGHT CURVES

An independent approach uses a parameterization of light curves, which have variable lengths, irregular sampling, etc., replacing them with a uniform set of statistical descriptors that can then be used to compose feature vectors, such as those described in [49]. Having the heterogeneously sampled light curves converted into a uniform set of feature vectors allows us to use a broad array of supervised and unsupervised classifiers, such as the ANN, SVM, Decision Trees (DT), etc. For our initial experiments we used DTs, where each internal node denotes a test on an attribute, each branch represents the outcome of the test and each leaf holds a class label.

We then search for an optimal set of features that give the best discrimination between different classes. We have applied a forward feature selection strategy that consists in selecting a subset of features from the training set that best predict the test data by sequentially selecting features until there is no improvement in prediction.

To avoid overfitting we use a 10-fold cross validation approach: the original sample is randomly partitioned into 10 subsamples. Each time a single subsample is retained as test, and the remaining are used as training data. This process is then repeated 10 times with each of the subsamples used exactly once as test. Table 1 shows the confusion matrix for the classification of 3 types of transient or variable sources from the CRTS, using 7 out of about 60 available features.

These are very encouraging results, and we continue to improve the feature selection. We note that different sets of features perform best for different class discriminations (see below). We will also be testing other supervised and unsupervised methods, in addition to DTs.

Table 1

| Class | Completeness | Contamination |
|---|---|---|
| Blazar | 83% | 13% |
| CV | 94% | 6% |
| RR Lyrae | 97% | 4% |

## V. A HIERARCHICAL APPROACH TO CLASSIFICATION

In the course of this work, it became obvious very quickly that different types of classifiers perform better for some event classes than for the others. Therefore, we took a hierarchical approach where some astrophysically motivated major features are used to separate different groups of classes, and then proceeding down the classification hierarchy using at every node those classifiers that are demonstrated to work best for that particular task, as illustrated schematically in Fig. 4.

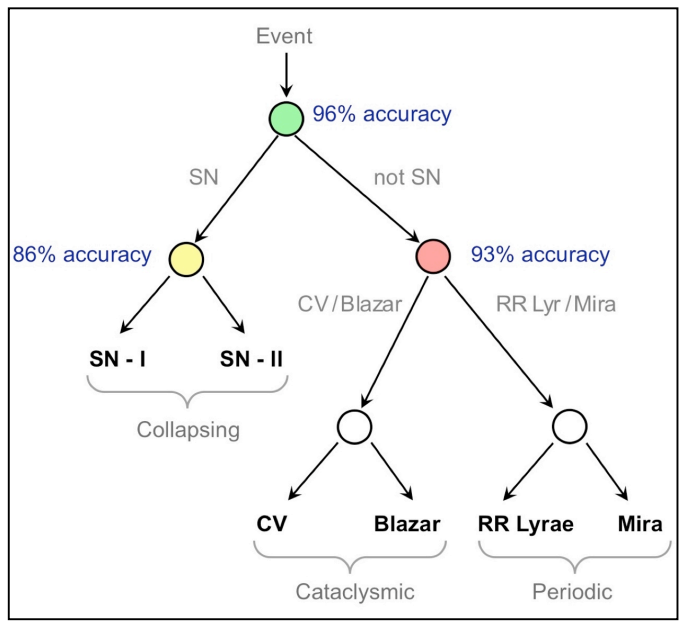

**Figure 4.** An approach to hierarchical classification in a decision tree mode. At each node we deploy an optimized combination of Bayesian classifiers to make a binary decision as indicated (the choices are informed by the prior astrophysical knowledge). We indicate the currently achieved best accuracies for the top 3 nodes; further work is in progress in optimizing the separation of the lower classification.

For example, events associated with Supernovae should have no prior outburst detections at the same location on the sky, since a star can only explode once, and repeated SN explosions from a small volume in any given galaxy would be exceedingly rare. Thus, if the past light curve shows any previous activity (a constant flux is OK, as it may be from the host galaxy), then the event cannot be a SN. Or, if a variable object shows variations between two flux bounds, it is very likely a pulsating variable star; if it shows more of an 1/f noise bursting behavior, then it could be either a cataclysmic variable star or a blazar; but if there is a radio source associated with it, then it is very likely a blazar; etc.

## VI. INCORPORATING CONTEXTUAL INFORMATION

Contextual information can be highly relevant to resolving competing interpretations: for example, the light curve and observed properties of a transient might be consistent with its being a cataclysmic variable star, a blazar, or a supernova. If it is subsequently found that that there is a galaxy in close proximity, the supernova interpretation becomes much more plausible. Such information, however, can be characterized by high uncertainty and absence, and by a rich structure – if there are two candidate host galaxies, their morphologies, distances, etc., become important, e.g., is this type of supernova more consistent with being in the extended halo of a large spiral galaxy or in close proximity to a faint dwarf galaxy? The ability to incorporate such contextual information in a quantifiable fashion is highly desirable. In a separate project we are investigating the use of crowdsourcing as a means of harvesting the human pattern recognition skills, especially in the context of capturing the relevant contextual information, and turning them into machine-processable algorithms.

A methodology employing contextual knowledge forms a natural extension to the logistic regression and classification methods mentioned above. Ideally such knowledge can be expressed in a manipulable fashion within a sound logical model, for example, it should be possible to state the rule that "a supernova has a stellar progenitor and will be substantially brighter than it by several order of magnitude" with some metric of certainty and infer the probabilities of observed data matching it. *Markov Logic Networks* (MLNs, [36]) are such a probabilistic framework using declarative statements (in the form of logical formulae) as atoms associated with real-valued weights expressing their strength. The higher the weight, the greater the difference in log probability between a world that satisfies the formula and one that does not, all other things being equal. In this way, it becomes possible to specify 'soft' rules that are likely to hold in the domain, but subject to exceptions - contextual relationships that are likely to hold such as supernovae may be associated with a nearby galaxy or objects closer to the Galactic plane may be stars.

A MLN defines a probability distribution over possible worlds with weights that can be learned generatively or discriminatively: it is a model for the conditional distribution of the set of query atoms $Y$ given the set of evidence atoms $X$. Inference consists of finding the most probable state of the world given some evidence or computing the probability that a formula holds given a MLN and set of constants, and possibly other formulae as evidence. Thus the likelihood of a transient being a supernova, depending on whether there was a nearby galaxy, can be determined.

The structure of a MLN – the set of formulae with their respective weights – is also not static but can be revised or extended with new formulae either learned from data or provided by third parties. In this way, new information can easily be incorporated. Continuous quantities, which form much of astronomical measurements, can also be easily handled with a hybrid MLN [37].

## VII. COMBINING AND UPDATING THE CLASSIFIERS

An essential task is to derive an optimal event classification, given inputs from a diverse set of classifiers such as those described above. A MLN approach could be used to represent a set of different classifiers and the inferred most probable state of the world from the MLN would then give the optimal classification. For example, a MLN could fuse the beliefs of different ML-based transient classifiers – four give a Supernova classification, and three give a Cataclysmic Variable, say – to give a definitive answer.

We are experimenting with the so-called "sleeping expert" [35] method. A set of different classifiers each generally works best with certain kinds of inputs. Activating these optionally only when those inputs are present provides an optimal solution to the fusion of these classifiers. Sleeping expert can be seen as a generalization of the *if-then* rule: *if* that condition is satisfied *then* activate this expert (in this case, a particular classifier), e.g., a "ML specialist" that makes a prediction only when the instance to be predicted falls within their area of expertise. For example, some classifiers work better when certain inputs are present, and some work only when certain inputs are present. It has been shown that this is a powerful way to decompose a complex classification problem. External or *a priori* knowledge can be used to awake or put experts to sleep and to modify online the weights associated with a given classifier; this contextual information may be expressed in text.

A crucial feature is the ability to update and revise the prior distributions on the basis of the actual performance, as we accumulate the true physical classifications of events, e.g., on the basis of follow-up spectroscopy. Learning, in the Bayesian view, is precisely the action of determining the probability models above – once determined, the overall model can be used to answer many relevant questions about the events. Analytically, we formulate this as determining unknown distributional parameters θ in parameterized versions of the conditional distributions above, $P(x \mid y = k; \theta)$. (Of course, the parameters depend on the object class $k$, but we suppress this below.) In a histogram representation, θ are just the probabilities associated with each bin, which may be determined by computing the histogram itself. In a Gaussian representation, θ would be the mean vector μ and covariance matrix Σ of a multivariate Gaussian distribution, and the parameter estimates are just the corresponding mean and covariance of the object-$k$ data. When enough data is available we can adopt a semi-parametric representation in which the distribution is a linear superposition of such Gaussian distributions. The corresponding parameters may be chosen by the Expectation-Maximization algorithm [13]. Alternatively, kernel density estimation could be used, with density values compiled into a lookup table [14,21].

We can identify three possible sources of information that can be used to find the unknown parameters, e.g., from the *a priori* knowledge, e.g. from physics or monotonicity considerations, or from examples that are labeled by experts, or from the feedback from the downstream observatories once labels are determined. The first case would serve to give an analytical form for the distribution, but the second two amount to the provision of labeled examples, $(x, y)$, which can be used to select a set of $k$ probability distributions.

## VIII. AUTOMATING FOLLOW-UP DECISION MAKING

We typically have sparse observations of a given object of interest, leading to classification ambiguities among several possible object types (e.g., when an event is roughly equally likely to belong to two or more possible object classes, or when the initial data are simply inadequate to generate a meaningful classification at all). Generally speaking, some of them would be of a greater scientific interest than others, and thus their follow-up observations would have a higher scientific return. Observational resources are scarce, and always have some cost function associated with them, so a key challenge is to determine the follow-up observations that are most useful for improving classification accuracy, and detect objects of scientific interest.

There are two parts to this challenge. First, what type of a follow-up measurement – given the *available* set of resources (e.g., only some telescopes/instruments may be available) – would yield the maximum information gain in a particular situation? And second, if the resources are finite and have a cost function associated with them (e.g., you can use only so many hours of the telescope time), when is the potential for an interesting discovery worth spending the resources?

We take an information-theoretic approach to this problem [15] that uses Shannon entropy to measure ambiguity in the current classification. We can compute the entropy drop offered by the available follow-up measurements – for example, the system may decide that obtaining an optical light curve with a particular temporal cadence would discriminate between a Supernova and a flaring blazar, or that a particular color measurement would discriminate between, say, a cataclysmic variable eruption and a gravitational microlensing event. A suitable prioritized request for the best follow-up observations would be sent to the appropriate robotic (or even human-operated) telescopes.

Note that the system is suggesting follow-up observations that may involve imperfect observations of a block of individual variables. This is a more powerful capability than rank-ordering individual variables regarding their helpfulness. Furthermore, we will ascertain that the framework accounts for the varying degrees of accuracy of different observations. The key to quantifying the classification uncertainty is the conditional entropy of the posterior distribution for object class $y$, given all the available data. Let $H[p]$ denote the Shannon entropy of the distribution $p$, which is always a distribution over object class $y$. (The classification is discrete, so we only need to compute entropies of discrete distributions.) Then, when we take an additional observation $x+$, uncertainty drops from $H[p(y \mid x0)]$ to $H[p(y \mid x0, x+)]$. We want to choose the observation $x+$ so that the expected final entropy is lowest.

Because all observing scenarios start out at the same entropy H [p(y | $x_0$)], maximizing entropy drop is the same as minimizing expected final entropy, $E[H[p(y | x_0, x_+)]]$. The expectation is with respect to the distribution of the new variable $x_+$, whose value is not yet known. Therefore, this entropy is a function of the *distribution* of $x_+$, but not the value of the random variable $x_+$. The distribution captures any imprecision and noise in the new observation. In our notation, the best follow-on observation thus minimizes, over available variables $x_+$,

$$H[p(y | x_+, x_0)] = -\sum_{y, x_+} p(y, x_+ | x_0) \log p(y | x_+, x_0).$$

This is equivalent to maximizing the conditional mutual information of $x_+$ about $y$, given $x_0$; that is, $I(y; x_+ | x_0)$ [22]. *The density above is known within the context of our assumed statistical model.* Thus, we can compute, within the context of the previously learned statistical model, a rank-ordered list of follow-on observations, which will lead to the most efficient use of resources.

Alternatively, instead of maximizing the classification accuracy, we consider a scenario where the algorithm chooses a set of events for follow-up and subsequent display to an astronomer. The astronomer then provides information on how interesting the observation is. The goal of the algorithm is to learn to choose follow-up observations which are considered most interesting, given the cost function constraints (e.g., the value of a limited amount of observing time at a given telescope or an instrument.

This problem can be naturally modeled using *Multi-Armed Bandit* algorithms (MABs) [38]. The MAB problem can abstractly be described as a slot machine with $k$ levers, each of which has different expected returns (unknown to the decision maker). The aim is to determine the best strategy to maximize returns. There are two extreme approaches: (1) exploitation - keep pulling the lever which, as per your current knowledge, returns most, and (2) exploration – experiment with different levers in order to gather information about the expected returns associated with each lever. They key challenge is to trade off exploration and exploitation. There are algorithms [47] guaranteed to determine the best choice as the number of available tries goes to infinity.

In this analogy different telescopes and instruments are the levers that can be pulled. Their ability to discriminate between object classes forms the returns. This works best when the priors are well assembled and a lot is already known about the type of object one is dealing with. But due to the heterogeneity of objects, and increasing depth leading to transients being detected at fainter levels, and more examples of relatively rarer subclasses coming to light, treating the follow-up telescopes as a MAB will provide a useful way to rapidly improve the classification and gather more diverse priors. An analogy could be that of a genetic algorithm which does not get stuck in a local maxima because of its ability to sample a larger part of the parameter space.

## IX. A Broader Applicability: Knowledge Discovery in Massive Data Streams

While our work was motivated largely by the acute and growing need for real-time data analysis and exploitation in astronomy and space science, the challenges we are tackling are common to many other fields, and it is easy to envision applications in fields as diverse as environmental monitoring, security, etc. In many situations, the intrinsically short time scales and large raw data volumes, combined with bandwidth limitations or signal latency, imply a need for a highly automated system; with machine learning, decision making, and rapid and prioritized follow-up response, without any human intervention. This has to be a dynamic process, incorporating the new data as they come in to make use of limited follow-up resources and constrained local processing capability (e.g., on a spacecraft, or in a field sensor). This presents a number of highly non-trivial challenges, some of which we addressed here, but with a potentially very broad applicability in terms of the methodology and technological solutions.

We are now dealing with Terascale data streams, and moving into the Petascale regime, but within a decade we will be facing challenges posed by data and event streams orders of magnitude larger and more complex, characterized by heterogeneity and diversity, in terms of both the phenomena, and the measurements. Tackling the problems of a rapid, automated classification and prioritization of interesting or anomalous events in massive data streams can only grow in relevance and a potential scientific utility.

More broadly, since the Petascale data sets and data streams exceed the human ability to inspect and explore them in any traditional way, incorporation of machine learning and machine intelligence tools is becoming a natural and a necessary part of the scientific method for the data-intensive research in the 21st century, a key component of the "fourth paradigm" [48].

### Acknowledgment

This work is supported in part by the NASA grant 08-AISR08-0085, the NSF grants AST-0909182 and IIS-1118041, by the W. M. Keck Institute for Space Studies, and by the U.S. Virtual Astronomical Observatory, itself supported by the NSF grant AST-0834235. Some of this work was assisted by the Caltech students Nihar Sharma and Yutong Chen, supported by the Caltech SURF program. We thank numerous collaborators and colleagues, especially within the CRTS survey team and the world-wide Virtual Observatory and astroinformatics community, for stimulating discussions.

*In Memoriam:* One of the authors (B. Moghaddam) tragically passed away after a brief illness, after this paper was submitted. We dedicate it to his memory. His contributions were important, and live on.